# Coherent dynamics in the rotor tip shear layer of utility scale wind turbines

**Xiaolei Yang**[1], **Jiarong Hong**[1,2], **Matthew Barone**[4] **and Fotis Sotiropoulos**[1,3]†

[1]St. Anthony Falls Laboratory, University of Minnesota, 2 Third Avenue SE, Minneapolis, MN 55414, USA

[2]Department of Mechanical Engineering, University of Minnesota, 111 Church Street SE, Minneapolis, MN 55455, USA

[3]Department of Civil, Environmental, and Geo- Engineering, University of Minnesota, 2500 Pillsbury Drive SE, Minneapolis, MN 55455, USA

[4]Sandia National Laboratories‡, Albuquerque, NM 87185 and Livermore, CA 94550, USA

Recent field experiments conducted in the near-wake (up to 0.5 rotor diameters downwind of the rotor) of a 2.5 MW wind turbine using snow-based super-large-scale particle image velocimetery (SLPIV) (Hong et al., Nature Comm., vol. 5, 2014, no. 4216) were successful in visualizing tip vortex cores as areas devoid of snowflakes. The so-visualized snow voids, however, suggested tip vortex cores of complex shape consisting of circular cores with distinct elongated comet-like tails. We employ large-eddy simulation (LES) to elucidate the structure and dynamics of the complex tip vortices identified experimentally. The LES is shown to reproduce vortex cores in good qualitative agreement with the SLPIV results, essentially capturing all vortex core patterns observed in the field in the tip shear layer. We show that the visualized vortex patterns are the result of energetic coherent dynamics in the rotor tip shear layer driven by interactions between the tip vortices and a second set of counter-rotating spiral vortices intertwined with the tip vortices. We further show that the mean flow within the region where such rich coherent dynamics occur satisfies the instability criterion proposed by Leibovich and Stewartson (J. Fluid Mech., vol. 126, 1983, pp. 335–356), indicating that the instability uncovered by the SLPIV and the LES is of centrifugal type. This study highlights the feasibility of employing snow voids to visualize tip vortices and demonstrates the enormous potential of integrating SLPIV with LES as a powerful tool for gaining novel insights into the wakes of utility scale wind turbines.

**Key words:**

## 1. Introduction

A novel super-large-scale particle image velocimetery (SLPIV) and visualization technique, which uses snowflakes as tracers, was recently developed by Hong *et al.* (2014) for measuring the near-wake turbulent flow structures of utility scale wind turbines. A field experiment using this technique was carried out on the Clipper Liberty 2.5 MW wind turbine at the EOLOS wind energy research field station (Hong *et al.* 2014; Chamorro *et al.* 2015*b*), University of Minnesota, from 2:00 am to 3:30 am on 22 February, 2013. The Clipper turbine has a rotor with diameter $D = 96$ m installed on a nacelle located 80 m above ground. The corresponding SLPIV setup is shown in figure 1(a). The tip vortex patterns visualized in this experiment at different time instants, which are identified as voids in the snow generated by centrifugal forces ejecting snow particles out of vortex cores, are shown in figure 1(b) to 1(g). We usually expect round-like tip vortex patterns as visualized in many model-scale experiments (Hu *et al.* 2012; Zhang *et al.* 2012; Sherry *et al.* 2013; Chamorro *et al.* 2015*a*). However, as seen in figure 1, vortex cores of complex shape were visualized resembling circular cores with attached elongated comet-like tails extending into the outer tip shear layer and inclined downwind. In addition to the pattern of a circular core with a single tail, which occurs most frequently in the visualization images, another pattern could also be identified resembling a core with two inclined tails like an inverted U-like shape (see figure 1(c)). As shown in figure 1(e), the angle of inclination of these void patterns appears largely around $33^o$. Moreover, patterns like those observed in figure 1 are found to persist in all the images recorded during our field deployment. These unusual (for wind turbine tip vortex cores) patterns of snow voids seem to indicate complex coherent dynamics near the outer tip shear layer the origin and governing mechanisms of which could not be identified using the SLPIV data. In this work, we employ high-fidelity large-eddy simulation with the turbine rotor modelled as rotating actuator surfaces to simulate the flow fields in the turbine near wake in order to elucidate the dynamics underlying these complex tip vortex patterns and uncover the governing mechanisms.

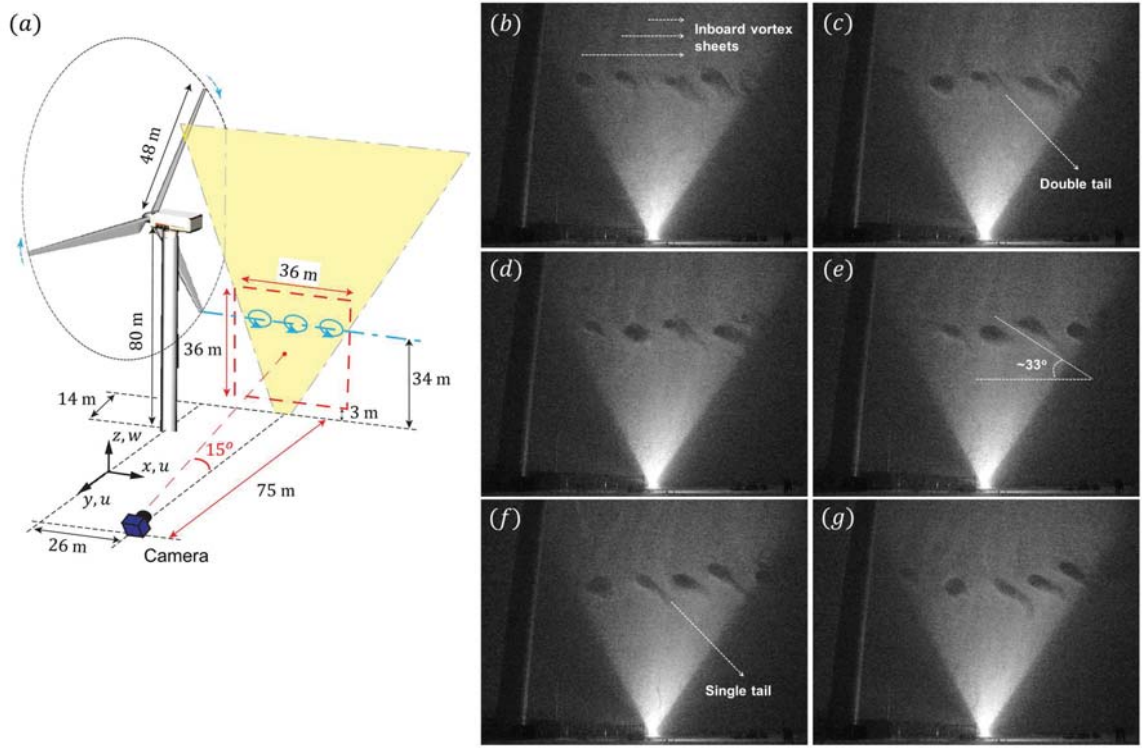

FIGURE 1. (a) Setup of SLPIV measurements adapted from Hong *et al.* (2014). (b)~(g) Snow voids at different time instants with time increment 1 s. The rotor period during the experiment is about 5 s. Circular voids with single and double tails are observed along the rotor tip shear layer. Vertical inboard vortex sheets are also visualized as lines of voids in the wake of the blades.

## 2. Numerical Methods

We carry out large-eddy simulation (LES) by solving numerically the spatially-filtered, unsteady, three-dimensional, incompressible Navier-Stokes and continuity equations. The subgrid scale stress tensor is modelled by the dynamic eddy-viscosity subgrid scale model (Germano *et al.* 1991). The governing equations are discretized on a hybrid staggered/non-staggered grid using three-point central finite differencing and integrated in time via a second-order accurate fractional step method. An algebraic multigrid acceleration along with GMRES solver is used to solve the pressure Poisson equation and matrix-free Newton-Krylov method is used for the filtered momentum equation. The numerical method satisfies the discrete continuity equation to machine zero at each time step. The details of the methodology can be found in Kang *et al.* (2011) and Yang *et al.* (2014).

We employ an actuator surface approach to model the rotor blades. The actuator surface concept was originally proposed by Shen *et al.* (2009) and allows for a better representation of the blade geometry than the standard actuator line approach. In the actuator surface approach, the forces imparted from the blade to the flow are distributed from the actuator surface formed by the airfoil chord instead of the actuator line (Sørensen & Shen 2002; Yang *et al.* 2014). The forces in the present actuator surface model are calculated based on a blade element approach instead from the pressure coefficient of the blades as in Shen *et al.* (2009). To distribute the force from the blade to the flow and interpolate the velocity to the blade location, we employ the discrete delta function in the form of $\delta_h(\boldsymbol{x}-\boldsymbol{X}) = \frac{1}{h^3}\phi\left(\frac{x-X}{h}\right)\phi\left(\frac{y-Y}{h}\right)\phi\left(\frac{z-Z}{h}\right)$, where $\boldsymbol{x}$, $\boldsymbol{X}$ are the coordinates of background grid and actuator surface grid, respectively, $\phi$ is the smoothed four-point cosine function proposed by Yang *et al.* (2009), with distribution width $5h$, and $h$ is defined below. In immersed boundary methods $h$ is equal to the grid spacing. In the actuator-type simulations of wind turbines, however, the width of the force distribution essentially determines the size of the tip vortex. This implies that the vortex core will become smaller if a finer mesh is employed. To eliminate this inherent grid dependency of the simulated tip vortices, here we select $h$ such that it can mimic the visualized by the SLPIV technique size of the tip vortex and satisfy the force conservation during the force distribution in the meantime. In the present case, the employed $h = D/80$ with the grid spacing near the turbine being equal to $D/160$. This results in the width of the force distribution to be $0.0625D$, which is consistent with the size of the snow voids measured in the field, i.e., approximately $0.042D$ to $0.073D$ ($4 \sim 7$ m).

## 3. Computational details

During the period of the SLPIV measurements, the wind blows from $54^o$ (clockwise from north) at 2:00 am to $34^o$ around 3:30 am. The measured velocity time series from the sonic anemometer installed at $z = 30$ m are employed to calculate the turbulence intensity characterizing the atmospheric turbulence around the rotor bottom tip during that time. For the data from 3:00 to 3:30 am on 22 February, 2013, the incoming wind speed $U_h$ on turbine hub height is about 5.67 ms$^{-1}$. The averaged wind direction is $35^o$ from the north. The turbulence intensity in the north-south, east-west and vertical directions are $0.108U_h$, $0.167U_h$ and $0.069U_h$. The thermal stratification is weakly stable. The tip-speed ratio based on the $U_h$ ranges from 8 to 11.

In the present large-eddy simulation, the size of the computational domain is $L_x \times L_y \times L_z = 4D \times 10.88D \times 5.21D$, in which $x$, $y$ and $z$ represent the streamwise, spanwise, vertical directions, respectively. The number of grid nodes is $N_x \times N_y \times N_z = 641 \times 434 \times$

384. The mesh is uniform around the turbine and in the streamwise direction, and is stretched in the spanwise and vertical directions at regions away from the turbine. The size of time step is $T/750$, where $T$ is rotational period of the turbine rotor. Free-slip boundary is applied at the top boundary and the spanwise boundaries. At the bottom boundary, a wall model based on the logarithmic law for calculating the shear stress on the ground is employed as follows

$$\frac{U}{u^*} = \frac{1}{\kappa} \log \frac{z}{z_0}, \quad (3.1)$$

where the Kármán constant $\kappa = 0.4$, the $z_0$ equals $5.83 \times 10^{-4} D$ in the present case. For the inflow at field scale, it is very difficult to reproduce via precursor simulations the actual atmospheric turbulence state with length scales spanning the range from the thickness of the atmospheric boundary layer ($\sim 1000$ m) down to the grid spacing resolving the turbine rotor (0.6 m in the present case) and taking into account the effects of thermal stratification. In the present work, the inlet flow is generated by superimposing the logarithmic mean profile (equation (3.1)) with the synthetic turbulence generated using the method by Mann (1998). No shear effect is considered in generating the synthetic turbulence, which means that the three components of the turbulence intensity are the same: $\sigma_u = \sigma_v = \sigma_w = 0.047 U_h$ for the present case. It is noted that the horizontal components of the turbulence intensity are significantly less than the corresponding measurements at 30 m. However this discrepancy will not have significant effects on the conclusions of this work since the objective of this study is not to compare the computed results with measurements quantitatively. The computation was first carried out until the total kinetic energy reached a quasi-steady state, and subsequently the flow fields were averaged for 45T.

In order to analyse the computed flow fields, the triple decomposition proposed by Hussain (1986) is employed, according to which an instantaneous quantity $f$ is decomposed as $f = \bar{f} + \tilde{f} + f'$, where $\bar{f}$ is the time-averaged contribution, $\tilde{f}$ is the coherent part of the fluctuations and $f'$ represents the incoherent fluctuating motion. The phase-averaged quantity $\langle f \rangle$ is computed by $\langle f \rangle = \bar{f} + \tilde{f}$.

## 4. Computational results

In this section we present qualitative comparisons between SLPIV and vorticity patterns obtained from LES and propose a physical mechanism that explains the field observations. To facilitate these comparisons it is important to re-iterate that SLPIV visualizes vortex cores in the form of voids in the snow. Centrifugal forces in the vortex cores eject snow particles out of the cores thus making the patterns of the vorticity field visible. Therefore, the specific void patterns that emerge should depend on the strength of the vortex cores, the underlying vorticity dynamics and the properties of the snow particles. Thus, a rigorous comparison of LES results with SLPIV should employ a Lagrangian analysis considering snow particle properties. Such an undertaking, however, is beyond the scope of the present work and will be left as subject for future research. Instead we will hypothesize that vortex strength and dynamics are dominant for determining the void patterns in SLPIV and employ contours of azimuthal vorticity as a surrogate to compare with the voids in the SLPIV images. An important point to keep in mind in this regard, is that voids in the snow will be created regardless of the sign of azimuthal vorticity as long as the vorticity magnitude is sufficiently large to eject snow particles out of the vortex cores. For that in figure 2(a) we plot the contours of the absolute value of the instantaneous azimuthal vorticity near the bottom tip at different time instants. It is readily apparent from this figure that the contours of azimuthal vorticity magnitude exhibit patterns that are very similar to the voids observed in the SLPIV experiment. Namely, we observe the same distinct rounded tip vortex cores along the tip shear layer with the elongated comet-like tails as in the field experiment. We also observe in this figure the vertical inboard vortex sheets that emanate from the turbine blades are also visible in the SLPIV images (see figure 1). Furthermore, it is worth noting that tip vortex cores with both single and double elongated tails appearing in figure 1 are observed in figure 2(a). In addition to these apparent similarities, it is also seen that the inclination angle of the tails in figure 2(a) is about $33^o$ which matches the field observation. To start unravelling the underlying coherent dynamics that give rise to these complex features, we plot in figure 2(b) contours of instantaneous azimuthal vorticity to clarify the sign of vorticity associated with the various void patterns. It is evident from these figures, that the vorticity tails in the LES are associated with both negative and positive vorticity. Negative (clockwise) azimuthal vorticity marks the tip vortex cores, which, however, exhibit thin tails of negative vorticity stretched away from the cores toward the outer tip shear layer and inclined forward. In between these tails of the tip vortices, elongated structures of positive azimuthal vorticity also appear, giving rise to a distinct pattern of counter-rotating vortical structures aligned along the outer edge of the turbine tip shear layer. This pattern remains coherent and persists up until $0.5D$ downwind (i.e. the window of observation of the SLPIV experiment) where it begins to break up into smaller-scale turbulent structures.

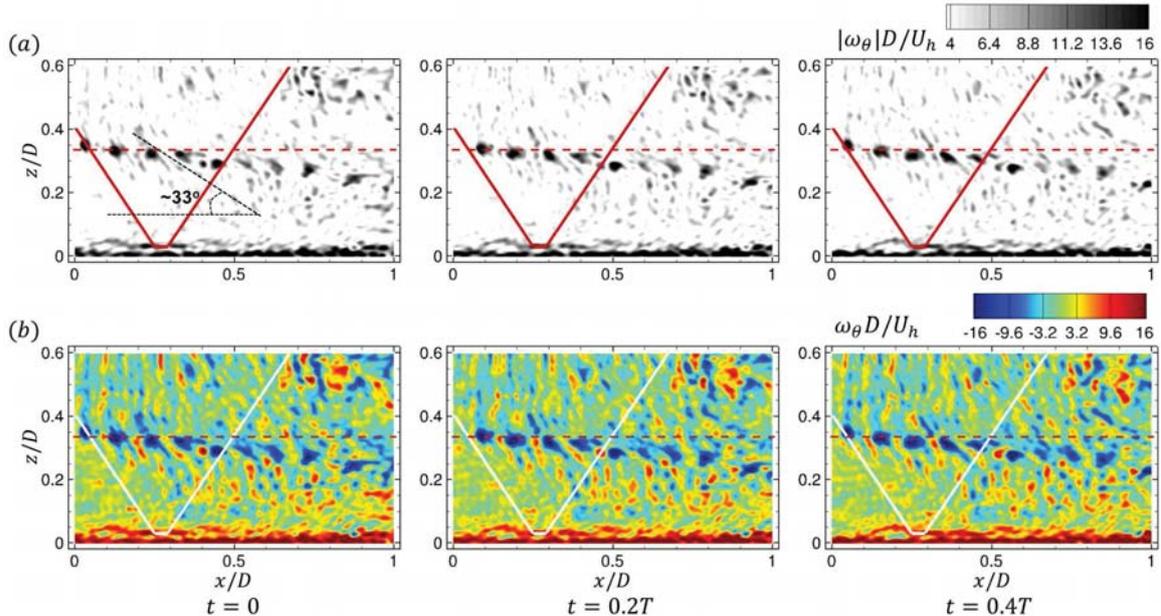

FIGURE 2. Contours of (a) the absolute value and (b) the corresponding actual value of the instantaneous azimuthal vorticity $\omega_\theta$ on a streamwise-vertical plane the same as in SLPIV at different time instants. $T$ is the period for one turbine rotation. Horizontal dashed line shows the elevation of the bottom tip of the rotor. Solid lines show the boundaries of the light in the SLPIV.

The above comparisons demonstrate the ability of the LES to capture vortex structures in good qualitative agreement with the SLPIV voids. They further show that the complex void patterns along the turbine tip shear layer are the result of pairs of counter-rotating vortices, namely the tip vortices with tail-like structures intertwined with elongated vortical structures of opposite sign. To further probe the structure of these counter-rotating vortices, we plot in figure 3(a) the contours of the phase-averaged azimuthal vorticity on the streamwise-vertical plane passing through the rotor center. As seen in

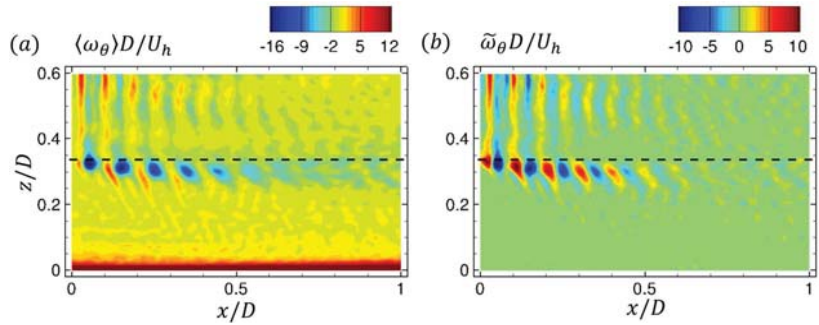

FIGURE 3. Contours of (a) phase-averaged azimuthal vorticity, and (b) the coherent part of the azimuthal vorticity on a streamwise-vertical plane passing through the rotor center. Horizontal black dashed line shows the elevation of rotor bottom tip.

figure 3(a), the blade tip vortices with the elongated tail-like structures and the counter-rotating vortices appear clearly in an alternating manner along the turbine tip shear layer up until $x/D < 0.5$ with the strength of the blade tip vortices being significantly larger than that of the counter-rotating vortices. Another important observation is that the phase-averaged tip vortices exhibit the distinct double-tail structure that was also observed in the instantaneous images in figure 2 and the SLPIV images (figure 1) starting at around $x = 0.15D$. Finally it is also evident form figure 3(a) that both the tip vortices and the counter-rotating vortices diminish in strength and are diffused rapidly for $x/D > 0.5$. To better elucidate the vortex structures near the tip, we plot in figure 3(b) the coherent part of the azimuthal vorticity ($\tilde{\omega} = \langle\omega\rangle - \bar{\omega}$), which visualizes the coherent dynamics along the tip shear layer. Figure 3(b) clearly shows that when the effect of mean shear is removed, the resulting counter-rotating vortex cores are similar both in size and magnitude. Moreover, the tails of the positive vortex cores become more pronounced and the negative (tip vortex) cores with double tails are also observed in the contours of $\tilde{\omega}$ between $x/D = 0.15$ and $x/D = 0.25$.

To investigate the impact of the rich dynamics we uncovered along the turbine tip shear layer on the turbulence statistics (the incoherent fluctuations in the triple decomposition), we plot in figure 4 contours of the turbulence kinetic energy (TKE), and the streamwise, spanwise (azimuthal) and vertical turbulence intensities ($\sigma_u$, $\sigma_v$ and $\sigma_w$). To facilitate the analysis of figure 4, we also superimpose the line contours of phase-averaged azimuthal vorticity. As seen in figure 4(a), significant increase in TKE is observed from the third blade tip vortex located between $x/D = 0.2$ and $x/D = 0.3$. The regions with high TKE are mainly located within the blade tip vortices (dashed contours near the bottom tip). Similar trends are also observed for the streamwise (figure 4(b)) and spanwise (azimuthal) (figure 4(c)) components of the turbulence intensities. It is worth noting though that the spanwise (azimuthal) intensity $\sigma_v$ has lower magnitude and it starts growing rapidly further downstream than $\sigma_u$. The vertical turbulence intensity contours, which characterize the outward stretching and transport of vortical structures along the tip shear layer, are seen in figure 4(d) to exhibit distinctly different trends than the horizontal and spanwise (azimuthal) counterparts. For instance, unlike $\sigma_u$ and $\sigma_v$, which show a monotonic increase starting at $x/D \sim 0.25$ and $x/D \sim 0.35$, respectively, $\sigma_w$ starts growing abruptly at $x/D \sim 0.15$, continuous to increase up until $x/D = 0.45$, but then starts declining gradually at further downwind locations. By examining the phase-averaged vorticity distributions in figure 3(a) and 3(b), we can see that the region of drastic change in the vertical turbulence intensity corresponds to the region where the tip and the counter-rotating spiral vortices co-exist and intertwine with each other. It is also observed in figure 4(d) that the pockets of high $\sigma_w$ intensity actually connect the cores of the blade tip vortices and the counter-rotating vortices. Overall the simulated patterns of vertical turbulence intensity presented herein reveal sudden increase in incoherent motions at approximately the location of the second tip vortex core ($x/D \sim 0.15$), which is then followed with continuous increase of the TKE especially in the area where the tip vortices lose coherence and dissipate ($x/D > 0.5$). This downwind increase in TKE, however, is driven by horizontal and spanwise (azimuthal) fluctuations. Vertical fluctuations are maximized in the region where the tip vortices engage in the complex interaction with the counter-rotating vortices ($0.15 < x/D < 0.5$), which results in ejections of vorticity toward the outer part of the shear layer in the form of tail-like, elongated structures.

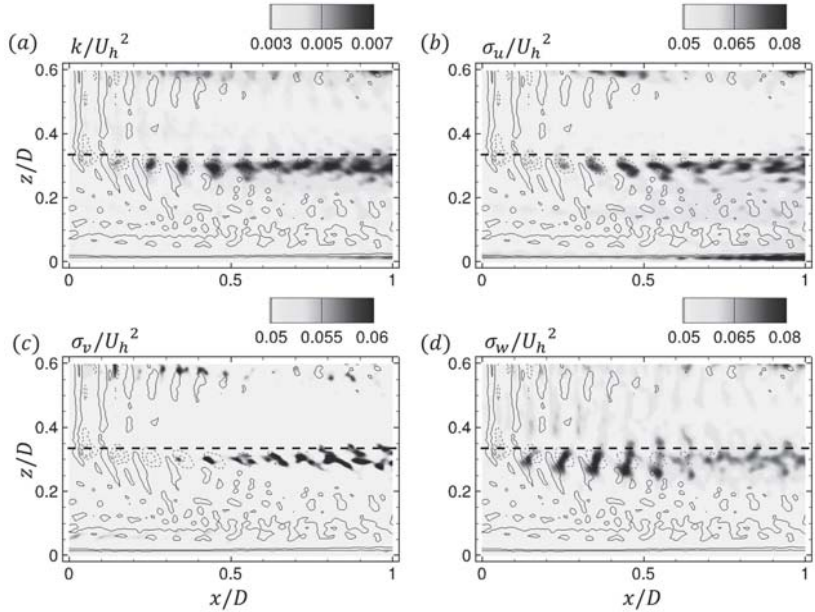

FIGURE 4. Flood contours of (a) turbulence kinetic energy, (b) streamwise turbulence intensity, (c) spanwise (azimuthal) turbulence intensity, and (d) vertical turbulence intensity superimposed with line contours of phase-averaged azimuthal vorticity (dashed: negative; solid: positive) on a streamwise-vertical plane passing through the rotor center. Horizontal black dashed line shows the location of the rotor bottom tip.

The results we have presented thus far show that the vortex cores with tails observed in the SLPIV images along the turbine tip shear layer are the result of rich and highly energetic coherent dynamics characterized by pairs of counter-rotating vortices. To elucidate the 3D structure of these vortices we plot in figure 5 instantaneous and phase-averaged iso-surfaces of azimuthal vorticity. The negative iso-surface of vorticity in figure 5 visualizes the tip vortices emanating from the blade tips. A second set of counter-rotating (positive) spiral vortices is also evident intertwined with the tip vortices in figure 5. Both sets of vortices lose coherence at approximately $0.5D$. It is also evident from figure 5(a) that the outer tip shear layer is characterized by intermittent ejections of positive and negative vorticity due to the interactions of the two sets of spiral vortices within the energetic tip shear layer. To our knowledge, this is the first time that such rich coherent dynamics is uncovered in the near wake of wind turbines–we note several previous studies (Okulov & Sørensen 2007; Ivanell *et al.* 2010; Iungo *et al.* 2013) that have focused exclusively on the far-wake instability of the tip vortices.

To identify a possible mechanism that leads to such complex coherent dynamics, we note that the flow in the wake of a wind turbine presents the potential for both shear and swirl induced instabilities. The former could arise as a result of both axial and azimuthal shear due to the axial velocity deficit and rotor-induced azimuthal velocity in the turbine wake, respectively. The rotor-induced swirl in the turbine wake, on the other hand, also raises the possibility of centrifugal instabilities. In fact examining the simulated radial profile of the mean circulation in the turbine near wake we find that it decreases with radius in the outer part of the rotor tip shear layer. Thus, the potential for centrifugal instability of the swirling turbine wake is certainly present at least in accordance to the classical theory by Rayleigh (1917) for purely swirling flows. A number of previous studies of related canonical flows have in fact demonstrated the potential for both shear (Kelvin-Helmholtz) and centrifugal instabilities to occur and compete with each other in swirling jets and wakes (Martin & Meiburg 1996; Loiseleux *et al.* 1998, 2000; Gallaire & Chomaz 2003). We note, however, that a major difference between the present flow and previously studied canonical swirling flows is that in the wind turbine wake case a specified number of vortices of a given sign are already imposed by the rotor in the form of the spiral tip vortices. As a result, the instability manifests itself in the form of second set of counter-rotating spiral vortices, which interact with the tip vortices and are stretched by the turbulent flow in the outer part of the tip shear layer giving rise to the distinct tail-like features observed in the snow visualization.

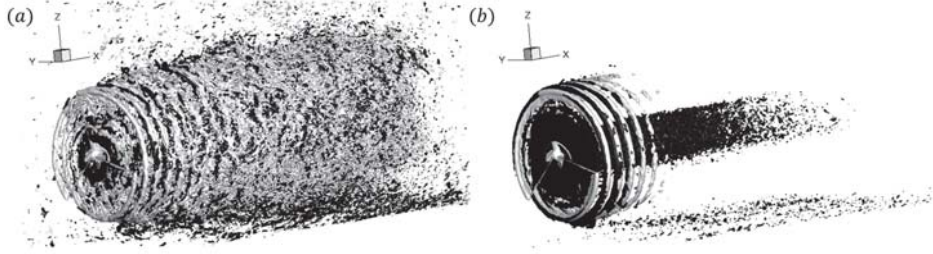

FIGURE 5. Isosurfaces of (a) instantaneous and (b) phased-averaged azimuthal vorticity $\omega_\theta$. In the region near the blade tip, the light isosurfaces show the tip vortices from the blades, while the dark isosurfaces show the counter-rotating spiral vortices. In (a), light and dark show the isosurfaces $\omega_\theta D/U_h$=-8.6 and 8.6, respectively. In (b), light and dark show the isosurfaces $\omega_\theta D/U_h$=-6.5 and 3.2, respectively.

To probe deeper into the nature of the instability we have uncovered herein, we examine the stability of the computed mean flow in the turbine wake using the criterion proposed by Leibovich & Stewartson (1983) for studying the stability of axisymmetric vortices with axial flow. This criterion assumes inviscid flow and ignores the streamwise variation of the mean velocity field. Both of these assumptions can be reasonable approximations of high Reynolds number flow in the near-wake of utility scale wind turbines, which rotate at high tip-speed ratios thus making the rotational component of the flow dominant in the vicinity of the turbine rotor. According to Leibovich & Stewartson (1983), an axisymmetric vortex with axial flow is centrifugally unstable if

$$C_{LS} = V_\theta \frac{d\Omega}{dr}\left[\frac{d\Omega}{dr}\frac{d\Gamma}{dr} + \left(\frac{dU}{dr}\right)^2\right] < 0, \quad (4.1)$$

where $V_\theta(r, x)$ is the time- and azimuthal-averaged azimuthal velocity, $U(r, x)$ is the time- and azimuthal-averaged axial velocity, $\Omega(r, x)$ is the time- and azimuthal-averaged angular velocity and $\Gamma(r, x)$ is the circulation $\int_0^{2\pi} rV_\theta d\theta$ ($\theta$ is the azimuthal angle). In figure 6(a), we plot the radial profiles of $C_{LS}$ at $0.1D$, $0.3D$, $0.5D$ and $0.7D$ downwind from the turbine. As seen, $C_{LS}$ becomes negative within the tip shear layer with the location of its minimum moving radially outward and its magnitude diminishing with distance from the turbine tip. In figure 6(b), we plot the flood contours of the coherent part of the azimuthal vorticity superimposed with line contours of $C_{LS}$. It is evident from this figure that the region within which $C_{LS}$ becomes negative and the instability criterion is satisfied coincides with the edge of the turbine tip shear layer and the location of the counter-rotating spiral vortices. These findings, therefore, make a strong case that the rich coherent dynamics we have uncovered in the near wake rotor tip shear layer is the manifestation of swirl-induced centrifugal instability. Furthermore, they point to the conclusion that the circular voids with comet-like tails identified by the SLPIV experiments arise by the interaction of the tip vortices with the counter-rotating spiral vortices in the centrifugally unstable turbine tip shear-layer, which are sheared and inclined forward by the outer flow.

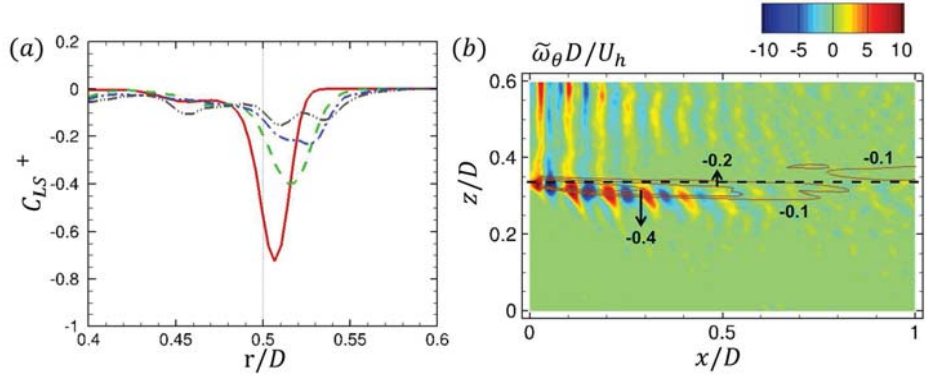

FIGURE 6. (a) Radial profiles of $C_{LS}$ at different downwind locations: $0.1D$ (solid line); $0.3D$ (dashed line); $0.5D$ (dash-dot line); $0.7D$ (dash-dot-dot line). The $C_{LS}$ is normalized using $D$ and $U_h$ as the characteristic length and velocity scales, respectively. (b) Flood contours of the coherent part of the azimuthal vorticity and line contours of the $C_{LS}$ (only the $C_{LS}$ contours near the bottom tip are plotted for clarity.) on a streamwise-vertical plane passing through the rotor center.

## 5. Summary

We have integrated field observation using snow visualization and SLPIV with LES to uncover and elucidate the physics of a new near-wake instability in the turbine tip shear-layer at utility scale. We showed that the turbine tip shear layer is unstable to centrifugal disturbances, which gives rise to counter-rotating spiral vortices intertwined with the rotor tip vortices. Future work should focus on studying the dependence of the onset and dynamics of this instability on the turbine tip speed ratio, inflow turbulence intensity, and atmospheric stability.

## Acknowledgments

This work was supported by Department of Energy DOE (DE-EE0002980, DE-EE0005482 and DE-AC04-94AL85000), Sandia National Laboratories and NSF Career Award (NSF-CBET-1454259) for Jiarong Hong. Computational resources were provided by Sandia National Laboratories and the University of Minnesota Supercomputing Institute.